\documentclass{article}
\usepackage{graphicx}
\usepackage{listings}
\usepackage{url}
\usepackage{tabularx}
\usepackage{amsmath}
\usepackage{hyperref} 

\newcommand{\footremember}[2]{%
    \footnote{#2}
    \newcounter{#1}
    \setcounter{#1}{\value{footnote}}%
}
\newcommand{\footrecall}[1]{%
    \footnotemark[\value{#1}]%
} 

\title{Reliability quality measures for recommender systems }

\author{
    Jesús Bobadilla \footremember{upm}{Universidad Politécnica de Madrid, Madrid, Spain}
    \and Abraham Gutierrez \footrecall{upm} 
    \and Fernando Ortega \footremember{u-tad}{U-tad: Centro Universitario de Tecnología y Arte Digital, Madrid, Spain}
    \and Bo Zhu \footremember{bit}{Beijing Institute of Technology, Beijing, China} 
}

\date{}

\begin{document}

\maketitle

\begin{abstract}
Users want to know the reliability of the recommendations; they do not accept high predictions if there is no reliability evidence. Recommender systems should provide reliability values associated with the predictions. Research into reliability measures requires the existence of simple, plausible and universal reliability quality measures. Research into recommender system quality measures has focused on accuracy. Moreover, novelty, serendipity and diversity have been studied; nevertheless there is an important lack of research into reliability/confidence quality measures.

This paper proposes a reliability quality prediction measure (\textit{RPI}) and a reliability quality recommendation measure (\textit{RRI}). Both quality measures are based on the hypothesis that the more suitable a reliability measure is, the better accuracy results it will provide when applied. These reliability quality measures show accuracy improvements when appropriated reliability values are associated with their predictions (i.e. high reliability values associated with correct predictions or low reliability values associated with incorrect predictions).

The proposed reliability quality metrics will lead to the design of brand new recommender system reliability measures. These measures could be applied to different matrix factorization techniques and to content-based, context-aware and social recommendation approaches. The recommender system reliability measures designed could be tested, compared and improved using the proposed reliability quality metrics.
\end{abstract}

\section{Introduction} \label{sec:intro}

In the Recommender Systems (RS) field, confidence or reliability has been defined as \cite{herlocker2004evaluating} ``How sure the recommender system is that its recommendation is accurate''. Quality metrics are a key factor for researchers in Collaborative Filtering (CF) RS. By combining quality measures (QM) and open datasets, researchers can improve results from previous works. RS researchers have focused on accuracy QM to test their methods and algorithms. Nevertheless, reliability measures (RM) and reliability quality measures (RQM) did not have the importance of accuracy or novelty research.

We claim RM are very important to RS users, since we know prediction and recommendation values have only a relative meaning. Electronic commerce clients usually look up the number of users that have rated products; we prefer a 4-star rated product based on 50 opinions to a 4.5 -star rated product based on 2 opinions. In this case, the client naive quality metric is just the number of opinions. Sometimes we check the mass function of opinions: we prefer a 3-star product based on twenty 3-star opinions to a 3-star product based on ten 1-star and ten 5-star opinions.

Following the above examples, an electronic commerce website could design a really simple RM combining both the number of ratings and the inverse of the rating's standard deviation. Additionally, it could add some other useful information such as the KNN number of neighbors involved in the prediction, content-based information, etc. This electronic commerce website could provide each client recommendation with the pair: 〈number of stars, reliability value). Clients would understand this information as a set of their friends recommending some films: ``we believe you will really love film A, but we are pretty sure you will like film B''; Film A $\langle 0.95,0.60\rangle$, Film B $\langle 0.70,0.92\rangle$.

\begin{figure}[ht]
\includegraphics[width=\textwidth]{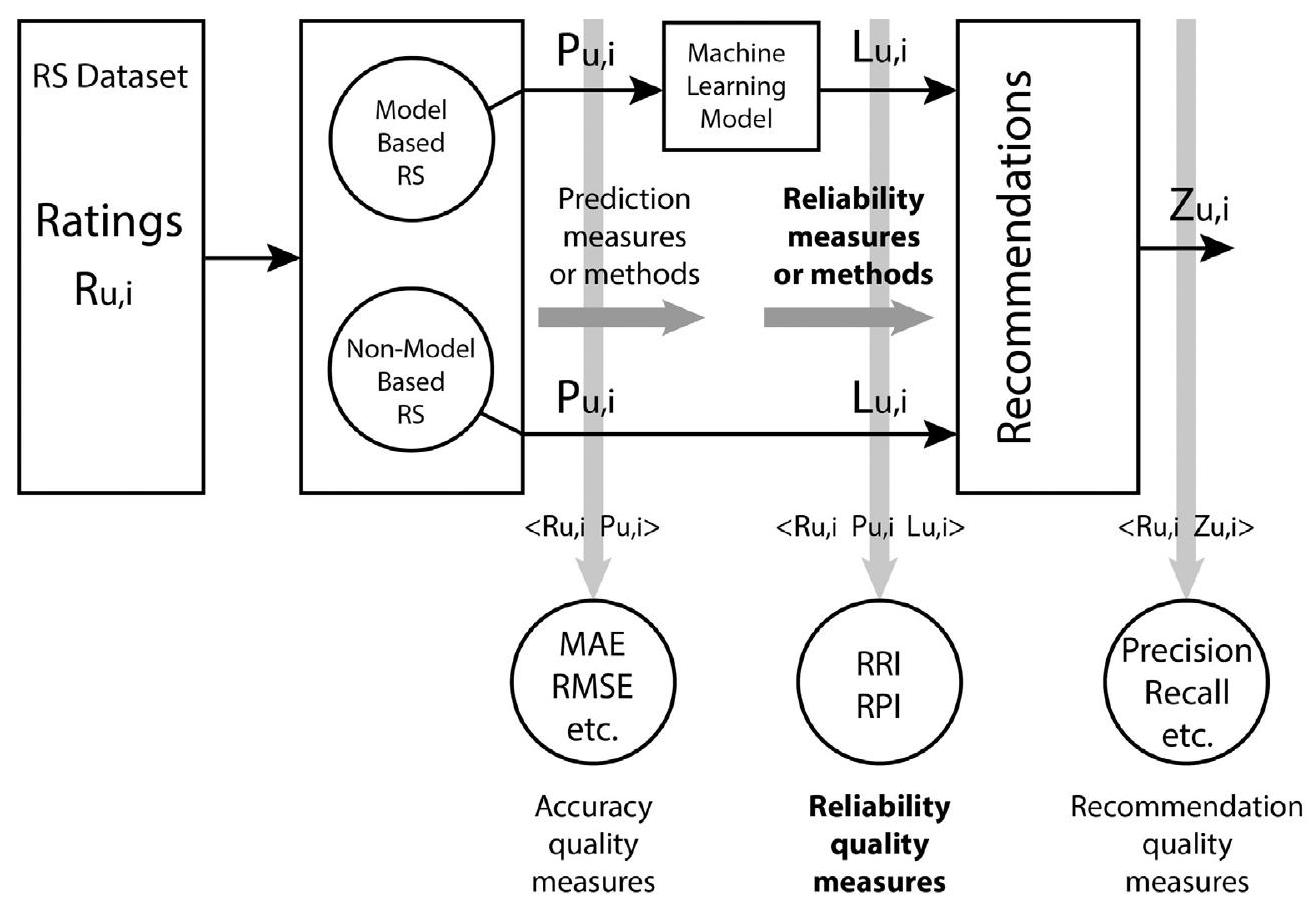}
\centering
\caption{Reliability measures and reliability quality measures.}
\label{fig:reliability-quality-measures}
\end{figure}

It is necessary to know the difference between a RM and a RQM. The first one assigns reliability values to each 〈user,item〉 prediction; the second one applies a testing strategy (such as cross-validation) to obtain the quality of the reliability values, that is the quality of the RM. Fig. \ref{fig:reliability-quality-measures} shows these concepts.

As can be seen in Fig. \ref{fig:reliability-quality-measures}, RM or reliability methods provide reliability values $(L u, i)$. These values can be obtained in the same RS stage that returns predictions, or they can be obtained using separate algorithms or methods. Using machine learning techniques (such as matrix factorization), reliability values could be obtained: a) directly from a modified machine learning algorithm, b) from the generated model (factorized matrices in the MF example). As far as we know there are no RS methods to get 〈prediction, reliability> pairs from machine-learning models only based on rating datasets; this is an important open research field that requires reliability quality metrics (such as the ones proposed) to be explored.

From Fig. \ref{fig:reliability-quality-measures} we can determine that testing reliability quality (vertical arrow) is different to obtaining reliability values (horizontal arrow). Whereas this paper focuses on testing RS reliability quality, obtaining reliability values from machine learning techniques or models will require specific research and will generate new publications.

Researchers can design a variety of RS-RM: a) based on the RS' nature and its data: content-based, collaborative, demographic, social, location-aware, etc. b) based on their filtering approaches: KNN, matrix factorization, bio-inspired methods, etc. It is necessary to test each proposed RM result, comparing each new approach with the existing ones. This process will lead us, as researchers, to improve the reliability values we provide to users, via their online service companies.

Testing RM requires some general, simple and suitable reliability quality metrics, e.g. for testing prediction and recommendation measures we generally use simple accuracy quality metrics: MAE, precision, recall, etc. The reliability quality metrics designed should be general enough to test reliability values coming from social filtering, KNN, matrix factorization methods, etc. They should be simple enough to be universally adopted and they should be based on an acceptable RS concept.

As far as we know, there is a lack of general purpose reliability CF-QM, such as the ones used to test accuracy or novelty: \cite{mazurowski2013estimating} ``while it appears to be acknowledged in the literature that an accurate estimation of prediction confidence would be of great use, little systematic research has been published toward this goal''. Early papers use the term ``confidence'' to define users' trust in the system recommendations \cite{herlocker2004evaluating,desrosiers2011recommender,swearingen2001beyond}. Later, some RM were established to assign confidence values associated to KNN \cite{bobadilla2010new,mclaughlin2004collaborative} parameters, \cite{hernando2013incorporating,moradi2015reliability} such as the neighborhood similarities to the active user.

Currently, the term ``reliability'' is often used. There are a variety of papers that incorporate social network trust relation RM to improve RS accuracy results \cite{park2016improving,martinez2015model}. CF reliability values have also been used to provide RS visual representations \cite{hernando2013trees,hernando2014hierarchical}. Finally, there are a few papers \cite{mazurowski2013estimating,hernando2013incorporating} that provide general purpose CF-RM and a tailored method to test their quality.

\begin{figure}[ht]
\includegraphics[width=\textwidth]{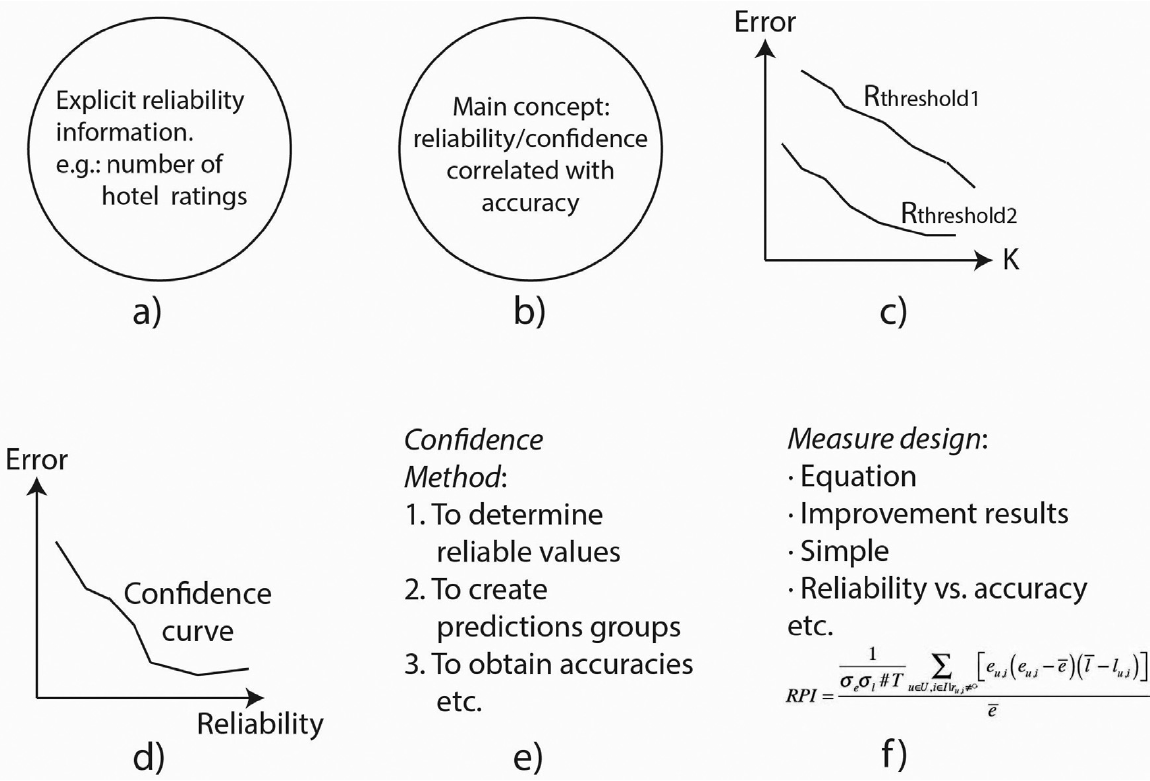}
\centering
\caption{Historical evolution on the reliability quality field (from a to e) and our proposal (f)}
\label{fig:reliability-evolution}
\end{figure}

In order to focus our proposal, we will briefly review existing solutions, and how they have been refined over the years. Fig. \ref{fig:reliability-evolution} shows the evolution experienced by the way of measuring RS reliability quality. Graphs a) to e) represent the state of the art, while graph f) outlines the proposal presented in this paper.

When suitable RQM methods did not exist, SR in operation provided users with explicit values that clients could use as reliability information. The most significant examples are the number of clients rating a product or the number of comments received by an item: Fig. \ref{fig:reliability-evolution}a.

From the year 2000, publications appeared that focused on the way of measuring RS quality \cite{herlocker2004evaluating,shani2011evaluating,swearingen2001beyond}. These publications focused on how to numerically achieve the accuracy of predictions and recommendations. They incorporate sections containing general concepts to measure other quality objectives ``beyond accuracy'', such as novelty, diversity, serendipity and confidence. Indications in these papers for RQM focus mainly on exposing a concept: values of confidence (reliability) and values of accuracy must be related, ``Probability that the predicted value is true'' \cite{shani2011evaluating}; Fig. \ref{fig:reliability-evolution}b.

Currently, two papers have appeared providing the quality of reliability approaches. One of these papers \cite{hernando2013incorporating} proposes a method to create CF-RM. To test RM qualities it establishes several reliability thresholds and, using graphs, it proves that ``the more reliable a prediction, the less liable it is to be wrong'' \cite{hernando2013incorporating}: Fig. \ref{fig:reliability-evolution}c. Some other experiments in paper \cite{hernando2013incorporating} present the concept of the ``confidence curve'': prediction reliability values are compared with prediction errors values. The expected behavior is an inverse relationship between reliability and error: Fig. \ref{fig:reliability-evolution}d. Mazurowski \cite{mazurowski2013estimating} shows a method to find the quality of the RM; this method is based on the confidence curve analysis: Fig. \ref{fig:reliability-evolution}e.

Our proposal (Fig. \ref{fig:reliability-evolution}f) is to provide a RQM. Unlike graphs, curves and methods, QM allows us to obtain quality values in a direct, simple and universal way, free of ambiguities and variations in implementation. We will use the method described in \cite{mazurowski2013estimating} as baseline that we will compare with our proposed RQM.

The next section, ``Related work'', selects the most significant references. Section \ref{sec:proposal}, ``Proposal'', explains the design considerations of the proposed RQM. Section \ref{sec:fundamentals}, ``Fundamentals'', formalizes the details of the RQM designed. Section \ref{sec:experiments}, ``Experiments'': a) introduces the scope and design of the paper's experiments, b) defines each RM tested, c) explains the baseline method, d) shows the quality results of both reliability predictions and reliability recommendations, and e) provides a discussion on the most significant results. Finally, Section \ref{sec:conclusions} provides the main conclusions of the paper and future works.

\section{Related work} \label{sec:rel-work}

In order to design a suitable reliability quality metric we need to understand the different approaches used to establish RM. Several reliability/confidence measures have been proposed, some of them focusing on specific areas: trust, social information, context-aware, etc. KNN based RM are often proposed.

The classic paper \cite{herlocker2004evaluating} differentiates the terms strength (prediction value) and confidence (reliability value) in a recommendation. They explain confidence as ``how sure the recommender system is that its recommendation is accurate''. This paper does not provide any specific confidence QM. The contribution made by our paper is to take the principles of the classic papers \cite{bobadilla2013recommender,herlocker2004evaluating,desrosiers2011recommender,shani2011evaluating,wu2012evaluating} and translate them into two formalized, unambiguous and easy to implement RS-RQM.

In \cite{herlocker2000explaining} the authors claim that suitable RS explanations of recommendations lead to increased user confidence. They show a simple confidence interval of 1-5 stars as an explanation for movie recommendations; they argue users can benefit from observing these confidence scores. From the user's point of view, ``RS preference ratings are malleable and can be significantly influenced by the recommendation received. The effect is sensitive to the perceived reliability of a RS and, thus, not a purely numerical or priming-based effect'' \cite{adomavicius2013recommender}. This approach is original, and has recently been expanded in \cite{hernando2013trees}. It shows processed information to the user, as reliability evidence. These papers \cite{adomavicius2013recommender,herlocker2000explaining,hernando2013trees} cover the model in Fig. \ref{fig:reliability-evolution}a. Our approach covers the alternative option: to calculate each reliability value from the rating matrix \cite{mazurowski2013estimating,hernando2013incorporating,moradi2015reliability,martinez2015model}. Reliability values can be used in different phases of RS; e.g. recommendation explanations, improving accuracy, etc.

McLaughlin and Herlocker \cite{mclaughlin2004collaborative} create a ``belief difference distribution'' indicating the belief in the prediction values. The intuition is that the less similar a neighbor is to the active user, the less belief we have that the neighbor's observed rating is the correct rating for the active user. This approach allows us to define a simple RM based on the KNN method. We provide an alternative, in which KNN-RM is taken as the variability of the neighborhood votes of each active user.

Shani and Gunawardana \cite{shani2011evaluating} define confidence in the recommendation as the system's trust in its recommendations or predictions. It shows the most common measurement of confidence to be the ``probability that the predicted value is true, or the interval around the predicted value where a predefined portion of the true values lie''. Our paper takes this same reasoning as a hypothesis. Finally, \cite{shani2011evaluating} found that users liked receiving some recommendations of items that they were already familiar with. They determined that to increase user confidence in the system, it is very important to find credible (known) recommendations.

Koren and Sill \cite{koren2011ordrec} predict a full probability distribution of the expected item ratings, rather than only a single score for an item. One of the advantages this approach brings is a novel method to estimate the confidence level in each individual prediction. Mazurowski \cite{mazurowski2013estimating} proposes several confidence (reliability) measures for individual rating predictions. They use the dataset ratings to extract confidence values associated to rating values. Mazurowski \cite{mazurowski2013estimating} also provides a method to obtain the quality of the RM. As far as we know, this method is the best RQM approach published, so we have adopted it as baseline for the RQM we propose in our paper. The reliability term is also used in the RS context as a QM to weight some other parameter, such as in \cite{bellogin2014neighbor}, where they introduce weights for neighbor selection. From the CF-KNN method, papers \cite{bellogin2014neighbor,koren2011ordrec,mclaughlin2004collaborative} implement probabilistic functions and weights to infer reliability values. Although their strategy is adequate and it obtains reasonable results, it suffers from generality because their approaches are KNN centered.

The reliability concept was used to weight the confidence of users and information. This concept has been currently used, in the RS field, to establish a RM to improve accuracy of trust-aware recommender systems \cite{moradi2015reliability,park2016improving,martinez2015model}. They assign reliability to users to improve the trust network. Trust and reliability have been combined in several papers. Often, this approach can only be applied to datasets incorporating additional information, such as content-based data or folksonomies. What we propose in this paper can be applied to any dataset. Hernando et al. \cite{hernando2013incorporating} provide a structured related work section focusing on trust, reputation, credibility and reliability/confidence terms in the CF area.

RM can be applied in the item to item or user to user CF process in order to get visual representations of RS databases \cite{hernando2013trees,hernando2014hierarchical}. User or item relations can be shown as edges of tree graphs; reliability of relations can be represented using different edge colors or transparencies. The RQM we propose can serve to determine which is the best RM for the visual representations proposed in \cite{hernando2013trees,hernando2014hierarchical}.

Hernando et al. \cite{hernando2013incorporating} define a general RM suitable for any arbitrary CF-RS. It also shows a method for obtaining specific RM specially suited to the needs of different specific RS. This approach effectively comprises the design of new CF-RM. Additionally, to test RM quality, they process the mean absolute error of the set of predictions having a reliability value higher than a threshold. This paper covers the graphs in Fig. \ref{fig:reliability-evolution}c and d.

\section{Proposal} \label{sec:proposal}

As shown in Fig. \ref{fig:reliability-evolution}f and explained in Section \ref{sec:intro}, we propose the design of RQM equations, which will improve the existing approaches: explicit information, graphs, confidence curves and methods. We work on the hypothesis that the more suitable a RM is the better accuracy results it provides when applied: predictions with higher reliabilities should provide more accurate (lower error) results, whereas we expect higher prediction errors on low reliability recommended items. This hypothesis follows the guidelines of the most relevant papers published in the area; by way of example, Shani and Gunawardana \cite{shani2011evaluating} state: a) ``perhaps the most common measurement of confidence is the probability that the predicted value is indeed true'', b) ``we can design for each specific confidence type a score that measures how close the method confidence estimate is to the true error in prediction'', and c) ``another application of confidence bounds is in filtering recommended items where the confidence in the predicted value is below some threshold''.

In paper \cite{mazurowski2013estimating}, the RQM is tested by analyzing the way in which reliability values are related to prediction errors. In particular, the ``confidence curve'' is evaluated. The confidence curve is represented by the reliability values on the $x$-axis and the prediction errors on the $y$-axis. \cite{hernando2013incorporating} is based on the same principle; it indicates: ``this RM is based on the usual notion that the more reliable a prediction the less liable it is to be wrong''. Finally, \cite{herlocker2004evaluating} states: ``evaluations of recommenders for this task must evaluate the success of high-confidence recommendations, and perhaps consider the opportunity costs of excessively low confidence''.

Along the lines of the papers reviewed, the design of the proposed RQM will follow the guidelines in Table \ref{tab:rqm-penalty-rewards} . We expect accurate RM to combine high reliability values with low prediction errors; this good behavior will be rewarded. On the contrary, high reliability values combined with high prediction errors will be penalized situations. High prediction errors are less serious if they are associated with low reliability values.

\begin{table}[ht]
\begin{tabular}{llll}
\hline
Prediction error & Reliability value & Result & Reliability quality \\
\hline
High & High & Big mistake & Big penalty \\
High & Low & Hit & Reward \\
Low & High & Big hit & Big reward \\
Low & Low & Mistake & Penalty \\
\hline
\end{tabular}
\caption{Reliability quality measure: penalty and reward situations.}
\label{tab:rqm-penalty-rewards}
\end{table}

Below are several principles and considerations that we have taken into account in the design of the RQM proposed in this paper:

\begin{itemize}
  \item To avoid non-formal mechanisms and also methods or algorithms: the results should be mathematical equations, which are easy to interpret and use, unambiguous and similar to the existing RS accuracy, novelty or diversity quality measures.
  \item To return quality improvement results, not absolute values. Results should show improvement or worsening of accuracy (error): improvement of results using reliability information, compared to results obtained without using that information. Thus, a ``0.15'' quality result indicates that we can improve the error by 15\% using the values of the RM. A ``-0.07'' quality result indicates that the error worsens by 7\%, and therefore the tested RM does not work at all.
  \item To provide two quality measures: 1) a measure of the reliability prediction quality, and 2) a measure of the reliability recommendation quality. The first one will be called RPI: ``Reliability Prediction Improvement''. The second one will be called RRI: ``Reliability Recommendation Improvement''.
  \item To avoid arbitrary parameters: using only ratings, predictions and reliability values. The only parameter that we can use is the threshold at which it is decided when a recommendation is relevant (as in the accuracy QM: precision and recall).
  \item To use only the ratings matrix, without relying on additional information that is not available in all RS datasets: social information, demographic data, context-aware raw data, etc.
  \item To encourage the stability of the results: processing the data in a uniform way, making use of all the data, avoiding saturation functions, discarding step functions, etc.
  \item To facilitate the integration of the QM into the current RS, providing simple, unambiguous, easy-to-interpret equations to implement and offering results that can be compared between different RM, collaborative filtering methods, and diverse datasets.
\end{itemize}

The current commercial RS can benefit from the proposed RQM in several ways:

\begin{itemize}
  \item Providing users with explicit and accurate reliability values: it is necessary to select the most appropriate RM to the RS dataset. Additionally, the chosen RM must fit with the nature and volume of the RS data. The proposed RQM perform the task of discovering the appropriate RM at each stage of the RS evolution. Explicit and accurate reliability values contribute to increase users' confidence.
  \item Providing users with implicit and accurate reliability values: in this case, the provided reliability information is non-explicit, e.g. ordering recommendations according to their reliability degree. As in the previous case, the proposed RQM is necessary to choose the correct RM.
  \item Improving prediction and recommendation accuracies: RQM allow to refine RM results, which make it possible to obtain accurate reliability values. One way to improve RS accuracy is to provide users with only the best predictions and recommendations that also have high reliability values.
  \item Providing to the RS administrators with an extra monitoring tool: RS administrators analyze, improve and fine tune their systems based on numerical parameters such as accuracy, reliability, coverage, etc. They need accurate quality measures, such as the MAE or the precision one. Currently there is no published RQM; this circumstance makes relevant the proposed RQM.
  \item Detecting particularly complex predictions and recommendations: e.g. cold start situations, non-standard users or items, shilling attacks, etc.
\end{itemize}

To incorporate this technology into a commercial RS, it must be taken into account that its execution is not carried out online, but rather a batch processing is done. Cross-validation methods, explained into Section 5, consume a large amount of execution time when applied to huge datasets. Reliability quality methods such as Mazurovsky's \cite{mazurowski2013estimating} fast resample consume too large amounts of execution time, while our proposed RQM is a lightweight process, based on only the execution of a metric. In summary, the incorporation of our RQM is completely feasible in commercial RS, due to its lightweight measure nature and to that it can be processed in parallel to the operation of the RS itself. Into Section 5, Fig. 4, we show and explain the implementation main steps of the cross-validation process tailored to our RQM.

The two equations proposed are formalized in the next section: Reliability Prediction Improvement (RPI) and Reliability Recommendation Improvement (RRI). Both RQM follow the principles and considerations described in the previous paragraphs.

\section{Fundamentals} \label{sec:fundamentals}
The RQM design strategy will penalize predictions simultaneously showing high accuracy errors and high reliability values. Predictions showing high accuracy errors and low reliability values are not penalized. The underlying principle is: ``the higher the prediction reliability value, the higher its accuracy error penalization; the lower the prediction error, the bigger its reliability value reward''.

Penalizations must be relative to the set of reliability values associated with the set of predictions or recommendations. In this way, if all the predictions reliability values were the same (i.e. 0.9 in [0..1]) we should not consider high and low values, and we should not apply an absolute penalization factor based on these values. This is the case for RS that do not provide reliability values: they can be considered as RS where their reliability parameters always have their maximum value (i.e. 1.0 in [0..1]). The accuracy of this type of RS will not be changed according to their implicit and identical reliability values.

The RQM design strategy must reward RM providing well balanced reliability density distributions, i.e. setting all the [0..1] reliability values within the subinterval [0.04..0.1] does not provide reliability variety to the users. A normal distribution (or similar) of the reliability values will provide a balanced and diverse set of reliabilities, where values higher than the mean distribution produce greater prediction error penalizations than values lower than the mean.

We define the following parameters (Table \ref{tab:parameters}).

\begin{table}[ht]
\begin{tabular}{ll}
\hline
$r_{u,i}$ & Rating of the user $u$ to the item $i$ \\
$p_{u,i}$ & Prediction of the rating $r_{u,i}$ \\
$l_{u,i}$ & Reliability associated with $p_{u,i}$ \\
$\bar{l}$ & Mean of the reliability values \\
$e_{u,i}$ & Error of $p_{u,i}$, $e_{u,i} = |p_{u,i}-r_{u,i}|$ \\
$\bar{e}$ & Mean of the error values \\ \\
\hline
\end{tabular}
\caption{Parameters.}
\label{tab:parameters}
\end{table}

To implement Table \ref{tab:rqm-penalty-rewards}, we will consider ``high'' reliability values to be the set of values greater than the mean reliabilities (and ``low'', to be the set of reliability values lower than the mean). Similarly, we will consider ``high'' prediction errors to be the set of values greater than the mean error predictions (``low'' to be the set of values lower than the mean). Table \ref{tab:rqm-penalty-rewards} can be redefined as:

\begin{table}[ht]
\begin{tabular}{llll}
\hline
Prediction error $\left(e_{u, i}-\bar{e}\right)$ & Reliability value $\left(\bar{l}-l_{u, i}\right)$ & Result & Reliability quality \\
\hline
Positive & Negative & Negative & Big penalty \\
Positive & Positive & Positive & Reward \\
Negative & Negative & Positive & Big reward \\
Negative & Positive & Negative & Penalty \\
\hline
\end{tabular}
\caption{Reliability quality measure: Penalty and reward situations.}
\label{tab:rqm-values}
\end{table}

High prediction errors in Table \ref{tab:rqm-values} (first column) correspond to errors of prediction greater than the mean absolute error (MAE): $e_{u, i}>\bar{e}$, and therefore, the expression $\left(e_{u, i}-\bar{e}\right)$ is positive; $\left(e_{u, i}-\bar{e}\right)$ is negative when prediction errors are lower than the MAE. Similarly, high reliability values in Table \ref{tab:rqm-values} (second column) correspond to reliability values greater than the mean reliability $\bar{l}: l_{u, i}>\bar{l}$, and therefore, the expression $\bar{l}-l_{u, i}$ is negative; $\bar{l}-l_{u, i}$ is positive when reliability values are lower than the reliability mean.

In order to achieve the results in Table \ref{tab:rqm-values} (third column) we combine prediction errors and reliability differences:

\begin{itemize}
  \item $\left(e_{u, i}-\bar{e}\right)\left(\bar{l}-l_{u, i}\right)$. We define $RPI_{u,i}=e_{u, i}\left(e_{u, i}-\bar{e}\right)\left(\bar{l}-l_{u, i}\right)$, as the ``Reliability Prediction Improvement'' related to the prediction of item $i$ to user $u$. A suitable reliability value $l_{u, i}$ associated with a prediction $\left\langle p_{u, i}, l_{u, i}\right\rangle$ will produce a positive value in $RPI_{u, i}$ wrong reliability values will produce negative values in $RPI_{u, i}$.
  \item RPI,``Reliability Prediction Improvement'' is the proposed RQM based on the MAE improvement. Eq. \ref{eq:rpi} shows its expression; it returns the average prediction improvement. RPI positive values reveal accurate RM, whereas RPI negative values reveal inappropriate ones. The higher the RPI result, the better the applied RM. The RPI equation denominator contains the traditional accuracy MAE, and therefore, RPI returns the MAE improvement value. The $1 /\left(\sigma_{e} \sigma_{l}|T|\right)$ term in Eq. (1) has the unique function of standardizing the RPI result. $T$ is the set of ratings, $U$ the set of users, and $I$ the set of items.
\end{itemize}

\begin{equation} \label{eq:rpi}
    \begin{aligned} 
        RPI & =\frac{\frac{1}{\sigma_{e} \sigma_{l} \# T} \sum_{u \in U, i \in I \mid r_{u, i} \neq .}\left[e_{u, i}\left(e_{u, i}-\bar{e}\right)\left(\bar{l}-l_{u, i}\right)\right]}{\bar{e}} \\
        where: \\
        T & =\left\{r_{u, i} \neq \cdot \mid u \in U \wedge i \in I\right\} \\
        e_{u, i} & =\left|p_{u, i}-r_{u, i}\right| \\
        \bar{e} & =M A E=\frac{1}{\# T} \sum_{u \in U, i \in I \mid r_{u, i} \neq \cdot} e_{u, i} \\
        \sigma_{e} & =\frac{1}{\# T} \sum_{u \in U, i \in I \mid r_{u, i} \neq \cdot}\left|\left(e_{u, i}-\bar{e}\right)\right| \\
        \sigma_{l} & =\frac{1}{\# T} \sum_{u \in U, i \in I \mid r_{u, i} \neq .}\left|\left(\bar{l}-l_{u, i}\right)\right|
    \end{aligned}
\end{equation}

Whereas RPI shows reliability prediction improvements, a second QM is proposed with the aim of dealing with ``Reliability Recommendation Improvements'': RRI. RPI has been based on the hypothesis that the more suitable a RM is, the better accuracy results it will provide when applied. RRI will be based on a similar hypothesis: the more suitable a RM is, the better recommendation results it will provide when applied. We expect to find that relevant recommendations are related to high reliability predictions.

We define:

\begin{itemize}
  \item $Z_{u}$ as the set of $N$ recommendations to user $u$.
  \item $\theta$ as the threshold to consider that a recommendation is relevant.
\end{itemize}

Both traditional precision and recall recommendation QM are based on the set of relevant recommendations (recommendations where $r_{u, i}>\theta$ ). Precision returns the proportion of relevant recommendations to the number of recommendations $(N)$; recall returns the proportion of relevant recommendations to the number of relevant items. The proposed RRI measure takes the precision and recall essential information (relevant recommendations) and tests the reliability values associated to each relevant recommendation.

Eq. \ref{eq:rel-recommendations} tests the reliability quality in all the relevant recommendations $\left(i \in Z_{u} \mid r_{u, i}>\theta\right)$ to all users $(u \in U)$. Relevant recommendations associated with high reliability values $\left(l_{u, i}>\bar{l}\right)$ produce positive results, whereas low reliability values $\left(l_{u, i}<\bar{l}\right)$ produce negative results. The higher the positive results the better the RM. Negative results show bad RM performance.

\begin{equation} \label{eq:rel-recommendations}
    \sum_{u \in U} \sum_{i \in Z_{u} \mid r_{u, i} \geq \theta}\left(l_{u, i}-\bar{l}\right)
\end{equation}

Eq. \ref{eq:rri} shows the RRI RQM for recommendations: it returns the gain factor between the quality of the relevant recommendations \cite{bellogin2014neighbor} and the total number of relevant recommendations. The higher the positive results the better the RM. The $1 / \sigma_{l}$ term in Eq. (3) has the function of standardizing the RRI result.

\begin{equation} \label{eq:rri}
    RRI=\frac{\frac{1}{\sigma_{l}} \sum_{u \in U} \sum_{i \in Z_{u} \mid r_{u, i} \geq \theta}\left(l_{u, i}-\bar{l}\right)}{\sum_{u \in U} \#\left\{i \in Z_{u} \mid r_{u . i} \geq \theta\right\}}
\end{equation}

The proposed RPI and RRI RQM can be applied to any type of RS (content -based, collaborative, location-aware, social, etc.) and to any RS method (KNN, matrix factorization, Bayesian, etc.), since they only need a set of pairs $<p_{u,i}, l_{u,i}>$ to operate. The $l_{u,i}$ values can be obtained using social network graphs, GPS coordinates, probabilistic values, etc.

\section{Experiments} \label{sec:experiments}

\subsection{Introduction}

In this section, the proposed prediction RQM (RPI) is tested as follows:

\begin{enumerate}
  \item Applying RPI to the results obtained by four RM: knn variability, support for user, support for item and fast resample. These RM are explained in the following subsection.
  \item Making use of two classic RS datasets: MovieLens 1M and the Netflix prize dataset.
  \item Comparing the RPI characteristics and results with the RQM published by Mazurovsky \cite{mazurowski2013estimating} (our baseline).

\end{enumerate}

As explained in the introduction section, Mazurovsky's method is the only one published that provides prediction RQM results testing several RM. Hernando et al. \cite{hernando2013incorporating} use a reliability threshold value to verify that above this value the accuracy results improve; this verification cannot be considered as a method or as a RQM. Consequently, the method devised by Mazurovsky is taken as baseline in this paper for the proposed prediction RQM. As far as we know there is no published recommendation RQM, therefore we cannot use a recommendation RQM baseline. We will test the recommendation RQM proposed in this paper (RRI) by applying it to the four selected RM.

\begin{figure}[ht]
\includegraphics[width=\textwidth]{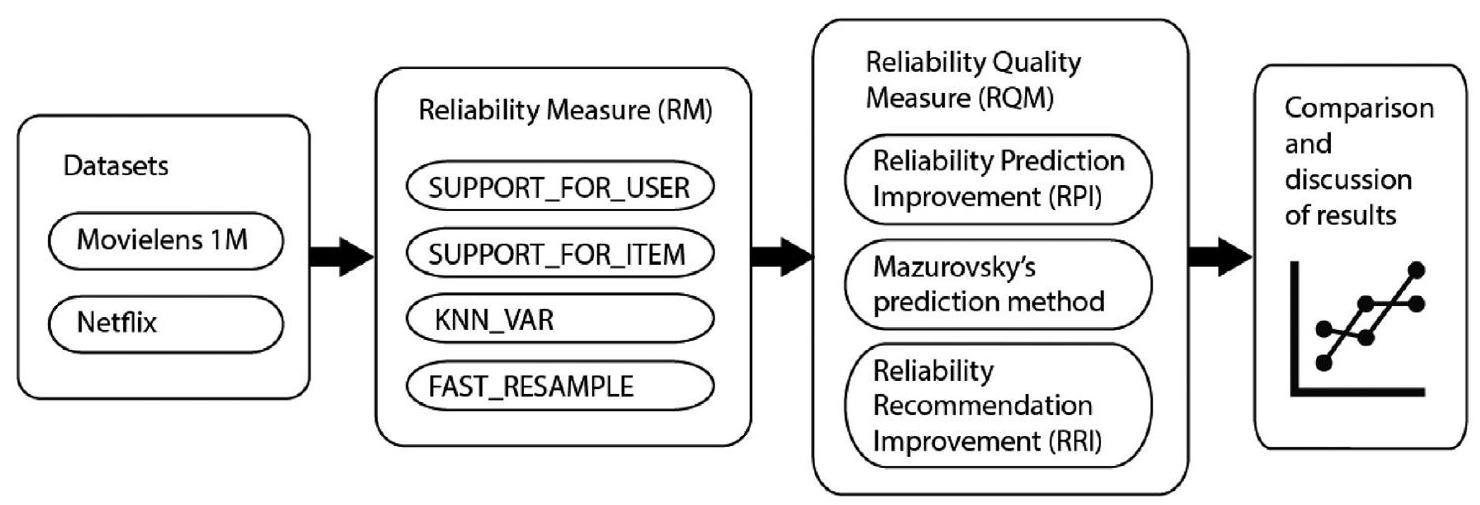}
\centering
\caption{Experimentation process stages.}
\label{fig:experiment-stages}
\end{figure}

Fig. \ref{fig:experiment-stages} shows a schematic of the experimentation process: 1) starting from two representative datasets in the RS field, 2) obtaining the prediction reliability values in each of the RM tested, 3) evaluating the RQM of each RM, and 4) comparing the quality of each RM and discussing the advantages and disadvantages of the proposed RQM (RPI) with regard to the baseline (Mazurovsky's method).

\begin{figure}[ht]
\includegraphics[width=\textwidth]{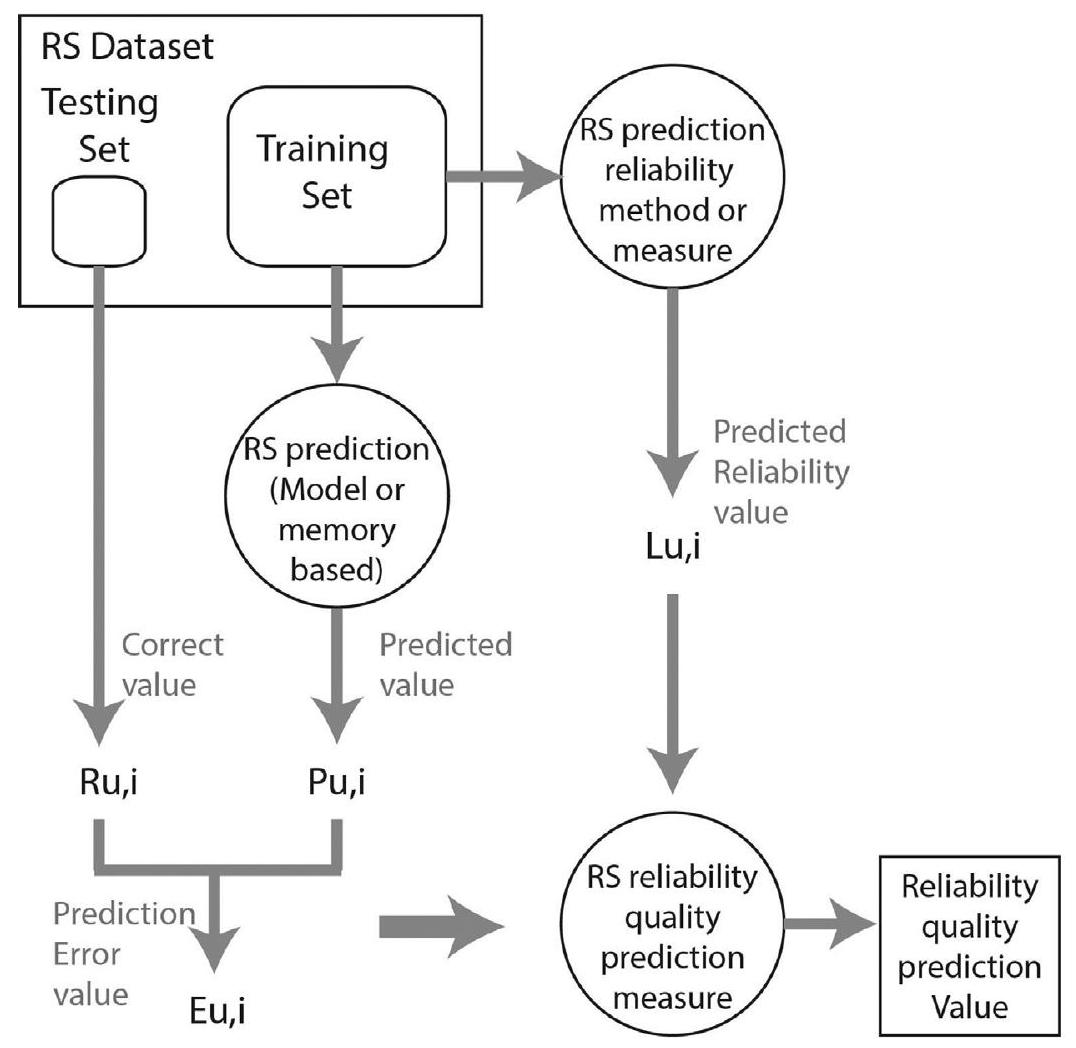}
\centering
\caption{Cross validation process to obtain reliability quality prediction values.}
\label{fig:cross-validation}
\end{figure}

Experiments were performed using 80\% training items and 20\% testing items, and the same proportions (80\% training, 20\% testing) with users. Test users and test items were taken at random from all users and items of each dataset. Fig. \ref{fig:cross-validation} shows the cross validation main steps to obtain reliability quality prediction values. We start from the disjoint RS training and testing sets. Testing set provides real ratings $R_{u, i}$ (correct values), whereas training set provides predicted values $P_{u, i}$ (applying some CF method). From correct ratings values and predicted ratings values, we obtain prediction error values $E_{u, i}$ (MAE). From the training set, the predicted reliability values $L_{u, i}$ are also obtained. Reliability quality prediction methods, measures and algorithms combine $R_{u, i}, P_{u, i}, E_{u, i}$ and $L_{u, i}$ to obtain the required reliability quality prediction values.

KNN prediction experiments were performed using a number of neighbors (K) from 20 to 400 , step 20 . KNN recommendation experiments were performed using $K=200$, a relevancy threshold $\theta=4$ and a number of recommendations $(N)$ from 2 to 20, step 2. The MovieLens $1 \mathrm{M}$ dataset was taken from \href{https://grouplens.org/datasets/movielens/}{https://grouplens.org/datasets/movielens/}. The Netflix prize dataset was collected between October 1998 and December 2005. The MovieLens $1 \mathrm{M}$ facts are: number of users: 4382, numbers of items: 3952 , number of ratings: $10,000,209$, min and max rating values: $1-5$. The Netflix facts are: number of users: 480,189 , numbers of items: 17,770 , number of ratings: $100,480,507$, min and max rating values: $1-5$. No preprocessing was carried out on the datasets.

The experiments section is structured as follows: 1) description of the tested RM, 2) explanation of the published method that we take as baseline in RQM prediction, 3) presentation of quality results in prediction reliability and comparing the proposed method with the baseline, 4) presentation of the quality results in recommendation reliability, and 5) discussion of the most relevant results.

\subsection{Tested reliability measures}
This section describes the operation of each RM that we are going to test. We chose the most significant RM from the Mazurovsky's paper \cite{mazurowski2013estimating}, and we added a RM not included in \cite{mazurowski2013estimating}. Mazurovsky tests the following RM: support for user, support for item, variability for item, resample, resample fast and inject noise. Resample and inject noise are non-scalable RM: the computational resources they require are too large, so in \cite{mazurowski2013estimating} resample fast is presented as the most suitable alternative to these two non-scalable RM. Variability for item is a simple RM, which we have replaced with an equivalent specialized RM: knn variability.

Table \ref{tab:tested-rqm} describes and formalizes each of the RM tested in the experiments. Note that RM have been ordered in Table \ref{tab:tested-rqm} from lowest to highest computational complexity.

\begin{table}[ht]
\footnotesize
\begin{tabularx}{\textwidth}{lXX}
\hline
Reliability Measure & Description & Formulation \\
\hline
Support for user & Number of ratings made by the user & $l_{u,i}=\#\{r_{u,i}|r_{u,i}\neq\bullet,j\in I\}$, \newline $\bullet$ means not voted  \\
Support for item & Number of ratings received by the item & $l_{u,i}=\#\{r_{s,i}|s_{u,i}\neq\bullet,s\in U\}$ \\
KNN variability & Inverse of the variance of the ratings made on the item by the user’s K-neighbors. & $l_{u,i}=\frac{\#V_{u,i}}{\sum_{s \in V_{u,i}} |r_{s,i}-\bar{v}_{u,i}}$, \newline $V_{u,i}=\{s \in K_u|r_{s,i} \neq \bullet\}$, \newline $\bar{v}_{u,i}=\frac{\sum_{s \in V_{u,i} r_{s,i}}}{\#v_{u,i}}$, \newline where $K_u$ is the set of neighbors of $u$ \\
Fast resample & RM Resample repeats the entire prediction process of the CF algorithm using different subsets of the R entire RS set of ratings. Fast resample uses a fraction of the original set to be resampled, making the process faster. For each of the N resampled matrices we obtain their set of predictions. Reliability is defined as the inverse of the predictions standard deviation. & $R$ = rating matrix, \newline $R^n$ = random select(R): \newline $|R^n|=\alpha R, \alpha \in [0..1], n \in [1..N]$, \newline $p^n$ = $R^n$ predictions, \newline $l_{u,i}=1/stdev\{p^n_{u,i}:n \in [1..N]\}$ \\
\hline
\end{tabularx}
\caption{Tested reliability measures.}
\label{tab:tested-rqm}
\end{table}

\subsection{Mazurovsky's method}

This section summarizes the method proposed in \cite{mazurowski2013estimating} to measure the quality of each RM tested. This method will act as baseline of the prediction RQM we present (RPI). Mazurovsky's method determines the relationship between the reliability values and their associated prediction errors. The steps of the method are:

\begin{enumerate}
  \item To determine a discrete set of $K$ reliability intervals (e.g. $k \in\{1..10\}$).
  \item To create groups of predictions. Each group $k$ contains the predictions with reliability $k$.
  \item To calculate the accuracy (MAE) of each prediction, obtaining $K$ groups of errors.
  \item To eliminate bigger errors (5\% of total errors).
  \item To measure the amplitude of the error interval (confidence interval) of each of the $K$ groups.
  \item To obtain the measure of quality from the variation of the interval sizes (confidence intervals) with respect to the variation of the values of $k$ (of reliability). This variation is expressed by a ``confidence curve''. Mazurowski \cite{mazurowski2013estimating} generically explains how this stage could be implemented. The specific solution they provide is to subtract the confidence curve values in $k=1$ and $k=K$.
\end{enumerate}

\begin{figure}[ht]
\includegraphics[width=\textwidth]{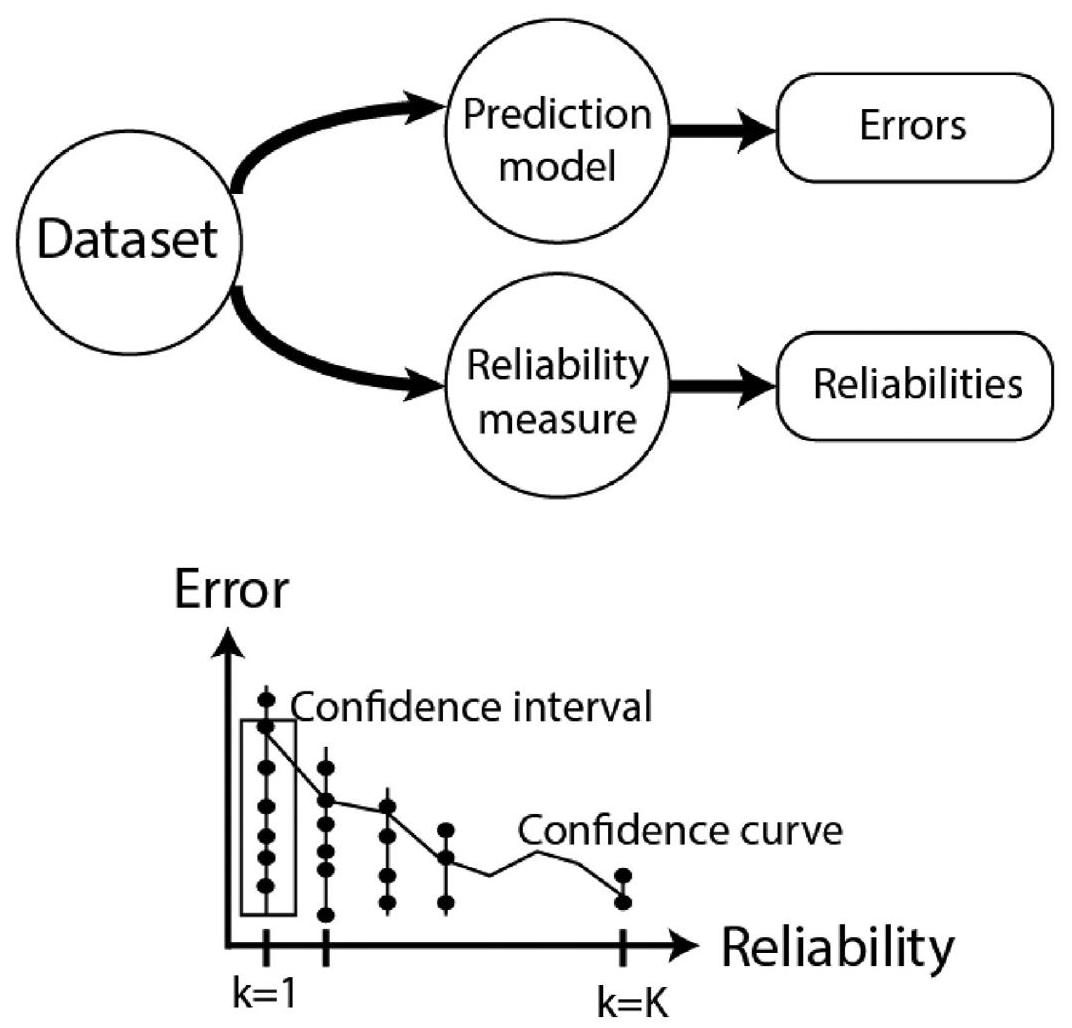}
\centering
\caption{Baseline method scheme.}
\label{fig:baseline-scheme}
\end{figure}

Fig. \ref{fig:baseline-scheme} graphically shows the elements involved in the baseline method.

\subsection{Prediction reliability quality results}

\begin{figure}[ht]
\includegraphics[width=\textwidth]{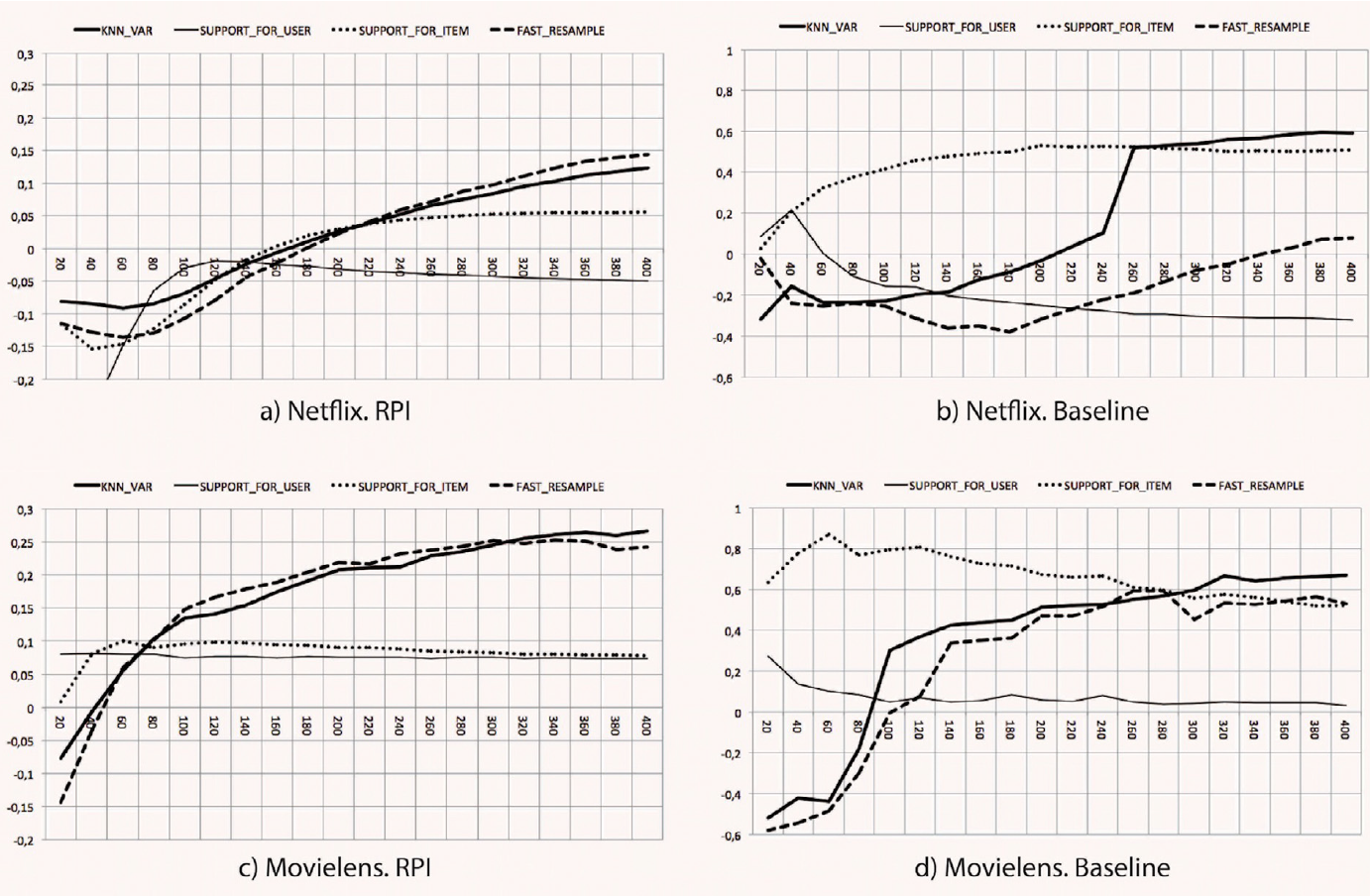}
\centering
\caption{Reliability quality results (referring to predictions) obtained when using several RM, reliability quality methods and recommender systems datasets. Axis $x$ : $K$ (number of neighbors), axis $y$ : error (a and c: error improvement; $b$ and $d$ : absolute error). Experiments parameters: cross validation: 20\% test users, 20\% test items.}
\label{fig:rpi}
\end{figure}

In this section we show the results of reliability quality on predictions: Fig. \ref{fig:rpi}. The two top graphs show the results making use of the Netflix dataset, while the two lower graphs show the results making use of the MovieLens $1 \mathrm{M}$ dataset. The graphs on the left show the results of the proposed method and the graphs on the right show the results obtained using the baseline \cite{mazurowski2013estimating}.

Graphs a) and c) from Fig. \ref{fig:rpi} show the superiority of RM fast resample and knn variability, followed by support for item in the case of the Netflix dataset. The results are consistent with the degree of complexity of RM: support for user and support for item are very simple and fast RM, but their results are not competitive, especially when applied to high values of $K$ (neighborhood).

Graphs a) and c) from Fig. \ref{fig:rpi} show that the results of the proposed RPI method are more stable and better explainable than their baseline \cite{mazurowski2013estimating} counterparts in Fig. \ref{fig:rpi} b) and d). The baseline method reports a quality hard to justify for the support for item RM. This situation can be produced due to the poor approximation assumed by the evolution of the confidence curve only according to its first and last values (step 6 in the ``Mazurovsky's method'' subsection). The discontinuities shown in graphs b) and d) are also, probably, due to the limitation indicated.

It is important to realize the difference in the scale of results of the proposed QM (RPI) and the baseline method: RPI reports the improvement (gain factor) that is reached when the correct predictions correspond to high reliabilities. The baseline method does not provide improvements: it returns the difference in the absolute value of the error between the maximum and the minimum reliability values. As an example: when using a value of $K=200$ in MovieLens $1 \mathrm{M}$ and $k n n$ variability, RPI indicates a $21 \%$ improvement, while the baseline method gives us an absolute half-point error gain.

By analyzing the scale of the $y$-axis on the graphs, we can verify that the improvements obtained are greater when using the MovieLens $1 \mathrm{M}$ dataset than when using Netflix: Increasing the size of the dataset, in general, also increases the reliability of its predictions, which reduces the margin of improvement of the RM: \cite{shani2011evaluating} ``the confidence in the predicted property also grows with the amount of data''.

Results obtained using the proposed method better explain the behavior of the RM than the results obtained using the baseline method: 1 ) the improvement of knn variability when the $K$ values increase is logical because the greater the number of neighbors the greater the margin to feed the variability tested by the RM, 2) the improvement of fast resample when the values of $K$ increase is also logical, because selecting a greater number of neighbors increases the impact of each resampling in the predictions obtained, and, therefore, increases the variation in the predictions of the different resamplings, and 3) the improvement of support for user and support for item, when increasing $K$, should not be as marked as the improvements obtained by knn variability, because in these cases the RM does not depend directly on the $K$ value.

In summary: 1) RPI produces results that adequately explain the logic behind RM, while the baseline method produces some results that do not fit the expected behavior, 2) RPI provides more balanced, continuous and homogeneous results than the baseline method, and 3) RPI offers relative values (of quality improvement), whereas the baseline method provides absolute values that cannot be easily compared and depend on the error scales of each dataset, on each collaborative filtering algorithm and on each RM tested.

\subsection{Recommendation reliability quality results}

In this section we show the results of reliability quality in recommendations: Fig. \ref{eq:rri}. Because there is no published measure or method that serves as baseline, only the recommendation RQM that we propose (RRI) is provided.

\begin{figure}[ht]
\includegraphics[width=0.6\textwidth]{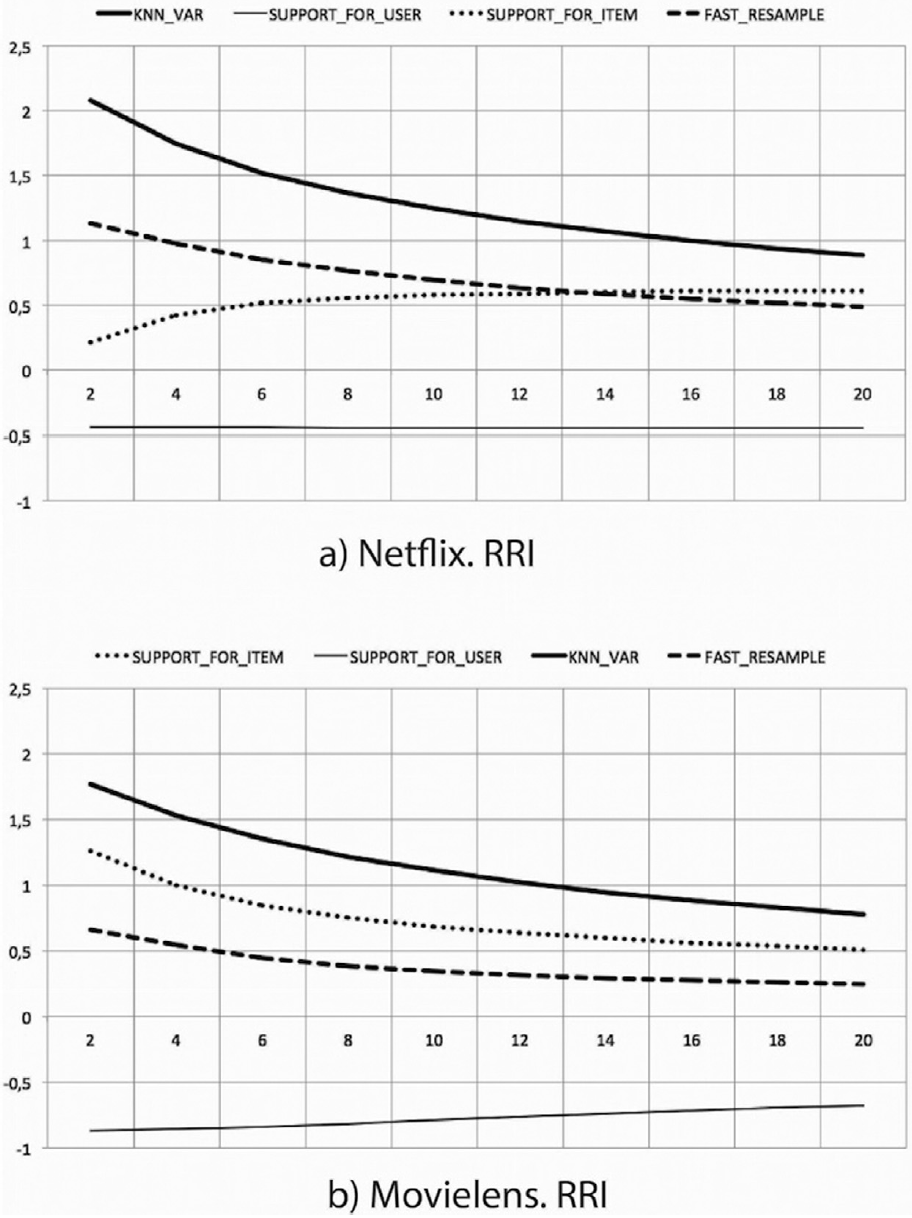}
\centering
\caption{Reliability quality results (referring to recommendations) obtained when using several reliability measures and recommender system datasets. Axis $x$ : $N$ (number of recommendations), axis $y$ : error improvement (gain factor). Experiment parameters: $K=200$, relevant threshold: 4 , cross validation: 20\% test users, 20\% test items.}
\label{fig:rri}
\end{figure}

Fig. \ref{eq:rri} shows the RRI improvements obtained on the tested RM: knn variability, fast resample, support for user and support for item. Both graphs (Netflix and MovieLens) present similar trends; in particular the trend to worsen quality is observed when the number of recommendations $N$ increases. This trend is logical, since the most promising recommendations ``disappear'' as we increase the number of items to recommend; e.g. if we have only 6 reliable and relevant items, then $N=6$ is adequate, but $N=14$ might not be, as the last eight recommendation reliability values will be lower and the risk of recommendation will be greater.

The results in Fig. \ref{eq:rri} present knn variability as the best-performing RM, while fast resample reduces the performance compared to their prediction quality results. The most plausible explanation is that fast resample is a RM based on stability. It is not based on CF parameters, as is the case of knn variability. In this way, the variation in the range of reliability values it provides is small, and it is not as effective to choose predictions with the highest values of reliability (recommendations).

By analyzing the $y$-axis scales in Figs. \ref{eq:rpi} and \ref{eq:rri}, it can be determined that recommendation improvements are much higher than prediction improvements, particularly when the number of recommendations is small. This difference is explained from a quantitative point of view: while prediction improvements are obtained by averaging a huge amount of predictions, the recommendation improvements refer to small amounts of recommendations. Thus, in the case of recommendations, only the relevant predictions that have very high values of reliability are chosen, and, therefore, improvements are higher.

General concepts:

\begin{itemize}
  \item Each RM provides different quality improvements, where knn variability shows the best performance, especially in the quality of recommendations.
  \item Fast resample and knn variability are RM suitable for measuring the quality of predictions.
  \item In general, the larger a RS dataset, the lower the improvement errors that will reach the QM.
  \item The recommendation error improvements are greater than the prediction improvements.
  \item The lower the number of recommendations, the greater the improvement errors given.
\end{itemize}

Comparison between the RQM proposed and the baseline method:

\begin{itemize}
  \item QM-RPI provides results that fit much better with those expected than the baseline method.
  \item RPI offers balanced and homogeneous results, while the baseline method generates variable and unstable results.
  \item RPI returns relative improvement results, whereas the baseline method returns absolute errors: RPI is more suitable to make comparisons and to establish analogies between QM, RM, datasets, and collaborative filtering methods.
  \item RPI is a QM defined by a single equation, whereas the baseline method is an algorithm: RPI is simpler, easier to understand, intuitive and universal.
  \item RPI lacks arbitrary parameters, whereas the baseline method requires some parameterized heuristics to interpret the confidence curve.
  \item As far as we know, no recommendation RQM approach has been published. We provide the recommendation QM-RRI.
\end{itemize}

\section{Conclusions} \label{sec:conclusions}

Users want to know the reliability of the recommendations; they look for the number of votes or comments on the films, hotels, products, etc. and they do not accept high predictions (usually 4 or 5 stars in e-commerce) if there is no reliability evidence.

Research into recommender system QM has focused on accuracy. Moreover, novelty, serendipity and diversity have been studied; nevertheless there is an important lack of research into reliability/confidence QM. It is important to promote RM associated with the predictions; the foundation for this research is to establish a set of appropriate RQM.

This paper proposes a reliability quality prediction measure (RPI) and a reliability quality recommendation measure (RRI). Both QMs are based on the hypothesis that the more suitable a RM, the better accuracy results it will provide when applied. These RQM show accuracy improvements when appropriate reliability values are associated with their predictions (i.e. high reliability values associated with correct predictions and low reliability values associated with incorrect predictions).

The results of the experiments performed indicate the superiority of the RM knn variability and fast resampling. They also show us that larger datasets have fewer opportunities for reliability improvements. In addition, the lower the number of recommendations we make, the better recommendation quality results we obtain.

The prediction RQM (RPI) proposed offers better characteristics than the existing baseline methods: 1 ) its results are in line with those expected, and they are also stable and progressive, 2) RPI results indicate improvements (gain factors), suitable for making comparisons between different RM and various datasets, 3) it is defined by an equation: it is simpler than a method or algorithm and does not contain arbitrary parameters.

This paper also provides a recommendation RQM: RRI. No recommendation RQM has been published in the RS field. Its experimental results show that the improvements are much higher in recommendations than in predictions, especially when the number of recommendations is low. This circumstance indicates the importance of: 1) making use of reliability values to calculate RS recommendations, and 2) more closely examining the development of new RM to be applied to recommendation tasks.

More promising future works will lead to the design of brand new recommender system RM. These measures could be applied to different matrix factorization techniques and to content-based, context-aware and social recommendation approaches. The recommender system RM designed can be tested, compared and improved using the proposed RQM.

\bibliographystyle{plain} 
\bibliography{references}

\begin{thebibliography}{10}

\bibitem{adomavicius2013recommender}
Gediminas Adomavicius, Jesse~C Bockstedt, Shawn~P Curley, and Jingjing Zhang.
\newblock Do recommender systems manipulate consumer preferences? a study of anchoring effects.
\newblock {\em Information Systems Research}, 24(4):956--975, 2013.

\bibitem{bellogin2014neighbor}
Alejandro Bellog{\'\i}n, Pablo Castells, and Iv{\'a}n Cantador.
\newblock Neighbor selection and weighting in user-based collaborative filtering: a performance prediction approach.
\newblock {\em ACM Transactions on the Web (TWEB)}, 8(2):1--30, 2014.

\bibitem{bobadilla2013recommender}
Jes{\'u}s Bobadilla, Fernando Ortega, Antonio Hernando, and Abraham Guti{\'e}rrez.
\newblock Recommender systems survey.
\newblock {\em Knowledge-based systems}, 46:109--132, 2013.

\bibitem{bobadilla2010new}
Jes{\'u}s Bobadilla, Francisco Serradilla, and Jesus Bernal.
\newblock A new collaborative filtering metric that improves the behavior of recommender systems.
\newblock {\em Knowledge-Based Systems}, 23(6):520--528, 2010.

\bibitem{desrosiers2011recommender}
Christian Desrosiers, George Karypis, F~Ricci, L~Rokach, B~Shapira, and PB~Kantor.
\newblock Recommender systems handbook.
\newblock {\em Recommender Systems Handbook. https://doi. org/10.1007/978-0-387-85820-3}, 2011.

\bibitem{herlocker2000explaining}
Jonathan~L Herlocker, Joseph~A Konstan, and John Riedl.
\newblock Explaining collaborative filtering recommendations.
\newblock In {\em Proceedings of the 2000 ACM conference on Computer supported cooperative work}, pages 241--250, 2000.

\bibitem{herlocker2004evaluating}
Jonathan~L Herlocker, Joseph~A Konstan, Loren~G Terveen, and John~T Riedl.
\newblock Evaluating collaborative filtering recommender systems.
\newblock {\em ACM Transactions on Information Systems (TOIS)}, 22(1):5--53, 2004.

\bibitem{hernando2013trees}
Antonio Hernando, Jes{\'u}s Bobadilla, Fernando Ortega, and Abraham Guti{\'e}Rrez.
\newblock Trees for explaining recommendations made through collaborative filtering.
\newblock {\em Information Sciences}, 239:1--17, 2013.

\bibitem{hernando2013incorporating}
Antonio Hernando, Jes{\'u}s Bobadilla, Fernando Ortega, and Jorge Tejedor.
\newblock Incorporating reliability measurements into the predictions of a recommender system.
\newblock {\em Information Sciences}, 218:1--16, 2013.

\bibitem{hernando2014hierarchical}
Antonio Hernando, Ricardo Moya, Fernando Ortega, and Jesus Bobadilla.
\newblock Hierarchical graph maps for visualization of collaborative recommender systems.
\newblock {\em Journal of Information Science}, 40(1):97--106, 2014.

\bibitem{koren2011ordrec}
Yehuda Koren and Joe Sill.
\newblock Ordrec: an ordinal model for predicting personalized item rating distributions.
\newblock In {\em Proceedings of the fifth ACM conference on Recommender systems}, pages 117--124, 2011.

\bibitem{martinez2015model}
Carmen Martinez-Cruz, Carlos Porcel, Juan Bernab{\'e}-Moreno, and Enrique Herrera-Viedma.
\newblock A model to represent users trust in recommender systems using ontologies and fuzzy linguistic modeling.
\newblock {\em Information Sciences}, 311:102--118, 2015.

\bibitem{mazurowski2013estimating}
Maciej~A Mazurowski.
\newblock Estimating confidence of individual rating predictions in collaborative filtering recommender systems.
\newblock {\em Expert Systems with Applications}, 40(10):3847--3857, 2013.

\bibitem{mclaughlin2004collaborative}
Matthew~R McLaughlin and Jonathan~L Herlocker.
\newblock A collaborative filtering algorithm and evaluation metric that accurately model the user experience.
\newblock In {\em Proceedings of the 27th annual international ACM SIGIR conference on Research and development in information retrieval}, pages 329--336, 2004.

\bibitem{moradi2015reliability}
Parham Moradi and Sajad Ahmadian.
\newblock A reliability-based recommendation method to improve trust-aware recommender systems.
\newblock {\em Expert Systems with Applications}, 42(21):7386--7398, 2015.

\bibitem{park2016improving}
Chanyoung Park, Donghyun Kim, Jinoh Oh, and Hwanjo Yu.
\newblock Improving top-k recommendation with truster and trustee relationship in user trust network.
\newblock {\em Information Sciences}, 374:100--114, 2016.

\bibitem{shani2011evaluating}
Guy Shani and Asela Gunawardana.
\newblock Evaluating recommendation systems.
\newblock {\em Recommender systems handbook}, pages 257--297, 2011.

\bibitem{swearingen2001beyond}
Kirsten Swearingen and Rashmi Sinha.
\newblock Beyond algorithms: An hci perspective on recommender systems.
\newblock In {\em ACM SIGIR 2001 workshop on recommender systems}, volume~13, pages 1--11, 2001.

\bibitem{wu2012evaluating}
Wen Wu, Liang He, and Jing Yang.
\newblock Evaluating recommender systems.
\newblock In {\em Seventh International Conference on Digital Information Management (ICDIM 2012)}, pages 56--61. IEEE, 2012.

\end{thebibliography}

\end{document}